\documentclass[12pt]{article}
\usepackage{hyperref}
\usepackage{amsmath,amsfonts,amssymb}
\usepackage{epsfig}
\usepackage{cite}
\usepackage{epstopdf}

\def\bra#1{\langle #1 |\, }
\def\ket#1{\, | #1 \rangle}
\newcommand{\ba}{\begin{array}}
\newcommand{\bat}{\begin{array}{cc}}
\newcommand{\ea}{\end{array}}
\newcommand{\no}{\nonumber}
\newcommand{\be}{\begin{equation}}
\newcommand{\ee}{\end{equation}}
\newcommand{\beqn}{\begin{eqnarray}}
\newcommand{\eeqn}{\end{eqnarray}}
\newcommand{\eqn}[1]{(\ref{#1})}
\newcommand{\cO}{{\cal O}}
\newcommand{\bel}[1]{\be\label{#1}}

\newcommand{\Leff}{L_{10}^{\mathrm{eff}}}
\newcommand{\Ceff}{C_{87}^{\mathrm{eff}}}

\usepackage{color}
\definecolor{orange}{RGB}{255,127,0}
\definecolor{brown}{RGB}{102,51,0}
\definecolor{myred}{RGB}{192,0,0}
\newcommand{\comment}[1]{}

\usepackage[normalem]{ulem}

\begin{document}
\bibliographystyle{unsrt}
\thispagestyle{empty}
\def\thefootnote{\fnsymbol{footnote}}

\begin{flushright}
IFIC/16-08\\  FTUV/16-0219
\end{flushright}

\vspace{1.5cm}

\begin{center}
{\fontsize{19}{23}\selectfont\bf
Updated determination of chiral couplings and \\[10pt]
vacuum condensates from hadronic $\boldsymbol{\tau}$ decay data}

\vspace{1.5cm}
Mart\'{\i}n Gonzalez-Alonso$,^{a}$
Antonio~Pich$^{b}$ and
Antonio Rodr\'{\i}guez-S\'anchez$^{b}$

\vspace*{.7cm}

$^a$
IPN de Lyon, CNRS and Universit\'e Lyon 1, F-69622 Villeurbanne, France

\vspace*{0.1cm}

$^b$
Departament de F\'\i sica Te\`orica, IFIC, Universitat de Val\`encia -- CSIC,\\
Apt. Correus 22085, E-46071 Val\`encia, Spain,

\vspace*{0.7cm}

\end{center}

\vspace*{1cm}

\begin{abstract}
We analyze the lowest spectral moments of the left-right two-point correlation function, using all known short-distance constraints and the recently updated ALEPH $V-A$ spectral function from $\tau$ decays. This information is used to determine the low-energy couplings $L_{10}$ and $C_{87}$ of chiral perturbation theory and the lowest-dimensional contributions to the Operator Product Expansion of the left-right correlator. A detailed statistical analysis is implemented to assess the theoretical uncertainties, including violations of quark-hadron duality.
\end{abstract}


\clearpage

\section{Introduction}

The hadronic decays of the $\tau$ lepton provide very valuable information on low-energy properties of the strong interaction, allowing us to analyze important perturbative and non-perturbative aspects of QCD \cite{Pich:2013lsa}. A very precise determination of the strong coupling can be extracted from the inclusive hadronic $\tau$ decay width \cite{Narison:1988ni,Braaten:1988hc,Braaten:1988ea,Braaten:1991qm,LeDiberder:1992te,LeDiberder:1992fr}, while the SU(3)-breaking corrections to the $\Delta S=1$ decay width \cite{Pich:1998yn,Pich:1999hc} are very sensitive to the Cabibbo quark mixing $|V_{us}|$ \cite{Gamiz:2002nu,Gamiz:2004ar}.
In this paper we are interested in the difference between the vector ($V$) and axial-vector ($A$) $\tau$ spectral functions, which gives a direct access to non-perturbative parameters related with the spontaneous chiral symmetry breaking of QCD \cite{Donoghue:1993xb,Davier:1998dz,Peris:2000tw,Bijnens:2001ps,Cirigliano:2003kc,Dominguez:2003dr,Narison:2004vz,Bordes:2005wv,Almasy:2006mu,Almasy:2008xu,GonzalezAlonso:2008rf,GonzalezAlonso:2010rn,GonzalezAlonso:2010xf,Boito:2012nt,Dominguez:2014fua,Boito:2015fra}.
A very detailed phenomenological study of the non-strange $V-A$ spectral function, using the 2005 release of the ALEPH $\tau$ data \cite{Schael:2005am}, was already done in Refs.~\cite{GonzalezAlonso:2008rf,GonzalezAlonso:2010rn,GonzalezAlonso:2010xf}.
The recent update of the ALEPH non-strange $\tau$ spectral functions~\cite{Davier:2013sfa} motivates an updated numerical analysis, based on the strategies developed in those references, which we present here. A comparison with other works available in the literature that employ different theoretical approaches will also be performed.

Compared to the 2005 ALEPH data set, the new public version of the ALEPH $\tau$ data incorporates an improved unfolding of the measured mass spectra from detector effects and corrects some problems \cite{Boito:2010fb} in the correlations between unfolded mass bins. The improved unfolding brings an increased statistical uncertainty near the edges of phase space. It has also reduced the number of bins in the spectral distribution, as a larger bin size has been adopted.

The starting point of our analysis is the two-point correlation function of the left-handed and right-handed quark currents:
\begin{align}
\label{eq:LRdef}
\Pi^{\mu\nu}_{ud,LR}(q)\; &\equiv\; i \int d^{4}x\, e^{iqx}\;
\bra{0}T\left(L^{\mu}_{ud}(x)R^{\nu \dagger}_{ud}(0)\right)\!\ket{0}
\nonumber\\
&=\; (-g^{\mu\nu}q^{2}+q^{\mu}q^{\nu})\;\Pi^{(1)}_{ud,LR}(q^{2})+q^{\mu}q^{\nu}\;
\Pi^{(0)}_{ud,LR}(q^{2})\, ,
\end{align}
where $L_{ud}^{\mu}(x)\equiv \bar{u}(x)\gamma^{\mu}(1-\gamma_{5})d(x)$ and $R_{ud}^{\mu}(x)\equiv \bar{u}(x)\gamma^{\mu}(1+\gamma_{5})d(x)$.
Owing to the chiral invariance of the massless QCD Lagrangian, this correlator vanishes identically to all orders in perturbation theory when $m_{u,d}=0$.
The non-zero value of
\beqn
\label{eq:PiLR}
\Pi(s) & \equiv &\Pi^{(0+1)}_{ud,LR}(s)\;\equiv\; \Pi^{(0)}_{ud,LR}(s)+\Pi^{(1)}_{ud,LR}(s)
=
\frac{2 f_\pi^2}{s-m_\pi^2} + \overline{\Pi}(s)
\eeqn
originates in the spontaneous breaking of chiral symmetry by the QCD vacuum, which results in different vector and axial-vector two-point functions. Thus, $\Pi(s)$ is a perfect theoretical laboratory to test non-perturbative effects of the strong interaction, without perturbative contaminations. The perturbative corrections induced by the non-zero quark masses are tiny and can be easily taken into account.
In Eq.~\eqn{eq:PiLR} we have made explicit the contribution of the pion pole to the longitudinal axial-vector two-point function. We will work in the isospin limit $m_u = m_d \equiv m_q$ where
the longitudinal part of the vector correlator vanishes.

\begin{figure}[tbh]\centering
\includegraphics[width=0.48\textwidth]{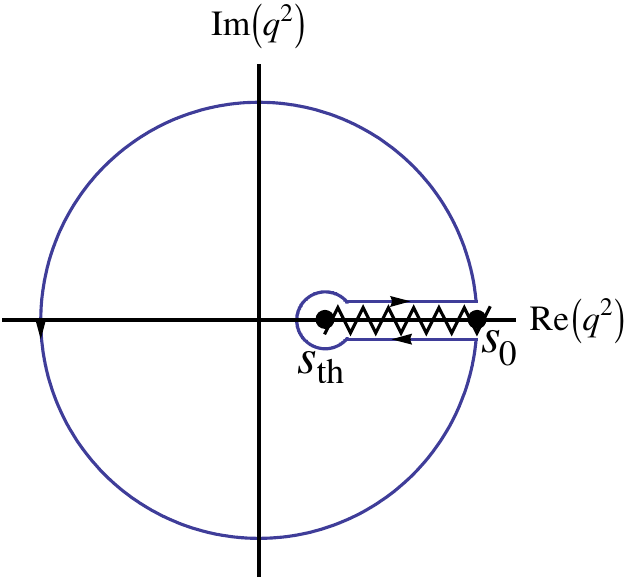}
\caption{Analytic structure of $\overline{\Pi}(s)$.}
\label{fig:circuit}
\end{figure}

The correlator $\overline{\Pi}(s)$ is analytic in the entire complex $s$ plane, except for a cut on the positive real axis that starts at the threshold $s_{\mathrm{th}} = 4 m_\pi^2$. Applying Cauchy's theorem in the circuit in Fig.~\ref{fig:circuit} to the function $\omega(s)\,\Pi(s)$, one gets the exact expression \cite{GonzalezAlonso:2010xf}:
\bel{eq:sumrule}
\int^{s_{0}}_{s_{\mathrm{th}}}ds\; \omega(s)\,\frac{1}{\pi}\operatorname{Im}\Pi(s)
+\frac{1}{2\pi i}\oint_{|s|=s_{0}}\! ds\; \omega(s) \,\Pi(s)\;
=\; 2\, f_{\pi}^{2}\,\omega(m_{\pi}^{2})+\operatorname{Res}[\omega(s)\Pi(s),s=0]\, ,
\ee
which relates the correlator in the Euclidean region, where it can be approximated by
its short distance Operator Product Expansion (OPE) \cite{Wilson:1969zs,Shifman:1978bx},
\bel{eq:OPE}
\Pi^{\mathrm{OPE}}(s)\; =\; \sum_{k}\frac{\mathcal{O}_{2k}}{(-s)^{k}}\; ,
\ee
with its imaginary part at Minkowskian momenta, accessible experimentally at low energies.
For $s_{th}\le s \le m_\tau^2$, the spectral function $\rho(s)\equiv\frac{1}{\pi}\operatorname{Im}\Pi(s)$ is determined by the difference between the vector and axial-vector hadronic spectral functions measured in $\tau$ decays.
Choosing different weight functions $\omega (s)$ one can change the sensitivity to different kinematical domains. We have only assumed that $\omega (s)$ is an arbitrary
analytic function in the whole complex plane except maybe at the origin where it can have poles, generating the corresponding residue $\operatorname{Res}[\omega(s)\Pi(s),s=0]$. The pion pole contribution is given by the term $2 f_{\pi}^{2}\,\omega(m_{\pi}^{2})$.

The OPE expresses the correlator as an expansion in inverse powers of momenta, which approximates very well $\Pi(s)$ in the complex plane, away from the real axis, at large values of $|s|$. Therefore, it provides a very reliable short-distance tool to compute the integral along the circle $|s|=s_0$, for sufficiently large values of $s_0$. The main source of uncertainty is the integration region near the real axis, but it can be suppressed with adequately chosen weight functions \cite{LeDiberder:1992fr}. In order to account for the small difference between the physical (exact) correlator and its OPE representation along the circle integration \cite{GonzalezAlonso:2010rn,Shifman:2000jv,Cirigliano:2002jy,Cirigliano:2003kc,Cata:2005zj}, one can introduce the correction \cite{GonzalezAlonso:2010rn,Cata:2005zj,Chibisov:1996wf}:
\begin{equation}\label{eq:DV}
\delta_{\mathrm{DV}}[\omega(s),s_{0}]\;\equiv\; \frac{1}{2\pi i} \oint_{|s|=s_{0}}ds\; \omega(s)\, \left[\Pi(s)-\Pi^{\mathrm{OPE}}(s)\right]
\; =\;\int^{\infty}_{s_{0}}ds\;\omega(s)\,\rho(s)\, ,
\end{equation}
which becomes zero at $s_0\to\infty$. A non-zero value of $\delta_{\mathrm{DV}}[\omega(s),s_{0}]$ signals a violation of quark-hadron duality in the spectral integration between $s_{\mathrm{th}}$ and $s_0$. We will discuss later the best strategy to control and minimize this kind of theoretical uncertainty.

Taking $\omega(s)=s^{n}$ with non-negative values of the integer power $n$, the pion pole is the only singularity within the contour. Therefore, the integral over the spectral function from $s_{\mathrm{th}}$ to $s_0$ is equal to the pion pole term $2 f_\pi^2 m_\pi^{2n}$, plus the OPE contribution $(-1)^n \cO_{2(n+1)}$ generated by the integration along the circle, up to duality violations (DV). However, in the chiral limit ($m_q=0$) and owing to the short-distance properties of QCD, $\Pi^{\mathrm{OPE}}(s)$ contains only power-suppressed terms from dimension $d = 2k$ operators, starting at $d = 6$ \cite{Bernard:1975cd}, which implies a vanishing OPE contribution for $n=0,1$:
\beqn
\int^{s_{0}}_{s_{\mathrm{th}}}ds\; \frac{1}{\pi}\operatorname{Im}\Pi(s)
 &=& 2f_{\pi}^{2} \, -\,\delta_{\mathrm{DV}}[1,s_{0}]\, ,
\label{wsr1}
\\[0.5cm]
\int^{s_{0}}_{s_{\mathrm{th}}}ds\; s\;\frac{1}{\pi}\operatorname{Im}\Pi(s)
 &=& 2f_{\pi}^{2}m_{\pi}^{2}\, -\,\delta_{\mathrm{DV}}[s,s_{0}]\, .
\label{wsr2}
\eeqn
The superconvergence properties of $\Pi(s)$ guarantee that the DV corrections to both sum rules approach zero very fast for increasing values of $s_0$.
When $s_0\to\infty$, there is no duality violation and one gets the well-known first and second Weinberg Sum Rules (WSRs) satisfied by the physical spectral functions \cite{Weinberg:1967kj}. With non-zero quark masses taken into account, the first relation is still exact, while the second gets a negligible correction of $\cO(m_q^2)$.

For higher values of the power $n$, Eq.~\eqn{eq:sumrule} gives relations involving the different OPE coefficients:
\bel{eq:On}
\int^{s_{0}}_{s_{\mathrm{th}}}ds\; s^n \; \frac{1}{\pi}\operatorname{Im}\Pi(s)
\; =\; (-1)^n\,\mathcal{O}_{2(n+1)}\, +\, 2 f_{\pi}^{2}m_{\pi}^{2n}
\, -\,\delta_{\mathrm{DV}}[s^{n},s_{0}]
\qquad\qquad (n\ge 2)\, .
\ee

For negative values of $n=-m < 0$, the OPE does not give any contribution
to the integration along the circle $s = s_0$, but there is a non-zero residue at the origen proportional to the $(m - 1)$th derivative of $\overline{\Pi}(s)$ at $s=0$.
At low values of $s$ the correlator can be rigorously calculated within chiral perturbation theory ($\chi$PT) \cite{Weinberg:1978kz,Gasser:1984gg,Gasser:1984ux,Ecker:1994gg,Pich:1995bw}. At present $\Pi(s)$ is known to $\cO(p^6)$ \cite{Amoros:1999dp}, in terms of the so-called chiral low-energy couplings (LECs) that we can determine through the relations:
\beqn
\int^{s_{0}}_{s_{\mathrm{th}}}ds\; s^{-1}\;\frac{1}{\pi}\operatorname{Im}\Pi(s)
&=& 2\,\frac{f_{\pi}^{2}}{m_{\pi}^{2}}\, +\, \Pi(0)\, -\, \delta_{\mathrm{DV}}[s^{-1},s_{0}]
\no\\& \equiv & -8\, \Leff\,
-\,\delta_{\mathrm{DV}}[s^{-1},s_{0}]\, ,
\label{eql10}\\[0.5cm]
\int^{s_{0}}_{s_{\mathrm{th}}}ds\; s^{-2}\;\frac{1}{\pi}\operatorname{Im}\Pi(s)
&=& 2\,\frac{f_{\pi}^{2}}{m_{\pi}^{4}}\, +\, \Pi'(0)
\, -\, \delta_{\mathrm{DV}}[s^{-2},s_{0}]
\no\\ &\equiv & 16\, \Ceff
\, -\,\delta_{\mathrm{DV}}[s^{-2},s_{0}]\, .
\label{eqc87}
\eeqn

The explicit expression of the correlator $\overline{\Pi}(s)$ at $\cO(p^6)$ in $\chi$PT is given in appendix~\ref{app:Pi}. The relation between the effective parameters  $\Leff$ and $\Ceff$ and their $\chi$PT counterparts, the LECs
$L_{10}$ and $C_{87}$, will be discussed in section~\ref{sec:ChPT}.

\section{A first estimation of the effective couplings}

\begin{figure}[tb]
\includegraphics[width=0.48\textwidth]{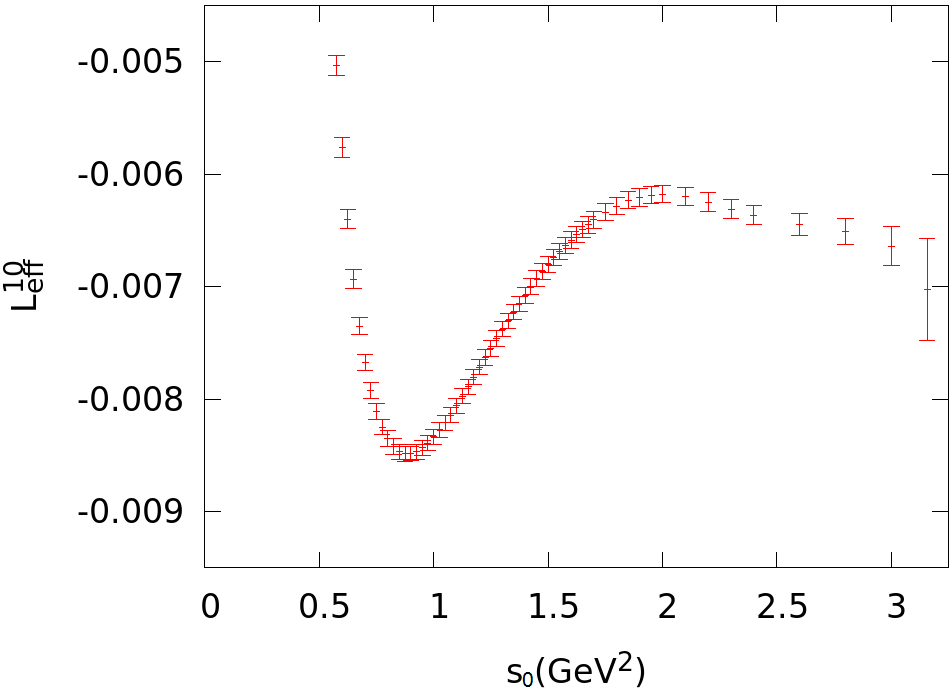}
\hfill
\includegraphics[width=0.48\textwidth]{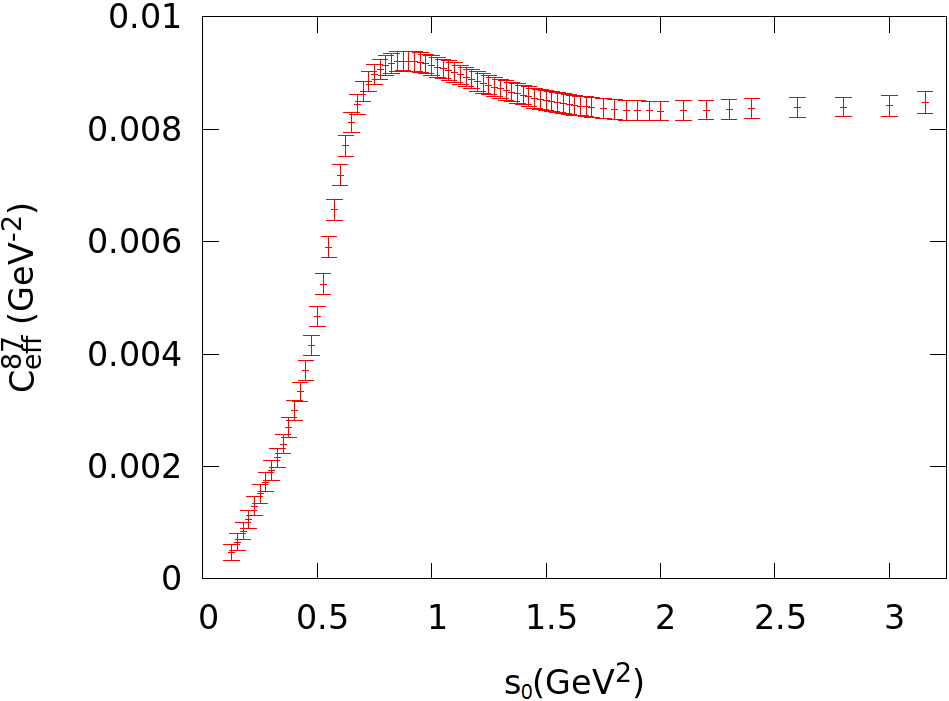}
\caption{$\Leff$ and $\Ceff$ from Eqs.~(\ref{eql10}) and (\ref{eqc87}), neglecting DVs, for different values of $s_0$.}
\label{fig.c87l101}
\end{figure}

Using the updated ALEPH spectral function \cite{Davier:2013sfa}, we can determine
$\Leff$ and $\Ceff$ with Eqs.~\eqn{eql10} and (\ref{eqc87}). As a first estimate, we neglect the DV terms and show in
Fig.~\ref{fig.c87l101} the resulting effective couplings, for different values of $s_{0}$. As expected and as it was already observed in Ref.~\cite{GonzalezAlonso:2008rf}, the results exhibit a strong dependence on $s_{0}$ at low energies, where the duality-violation corrections are not negligible. At larger momentum transfers the curves start to stabilise, indicating that the violations of duality become smaller. However, especially for $\Leff$, the curves are not yet horizontal lines at $s_{0}$ near $m_{\tau}^{2}$, which implies that duality-violation effects are still present.

\begin{figure}[tb]
\includegraphics[width=0.48\textwidth]{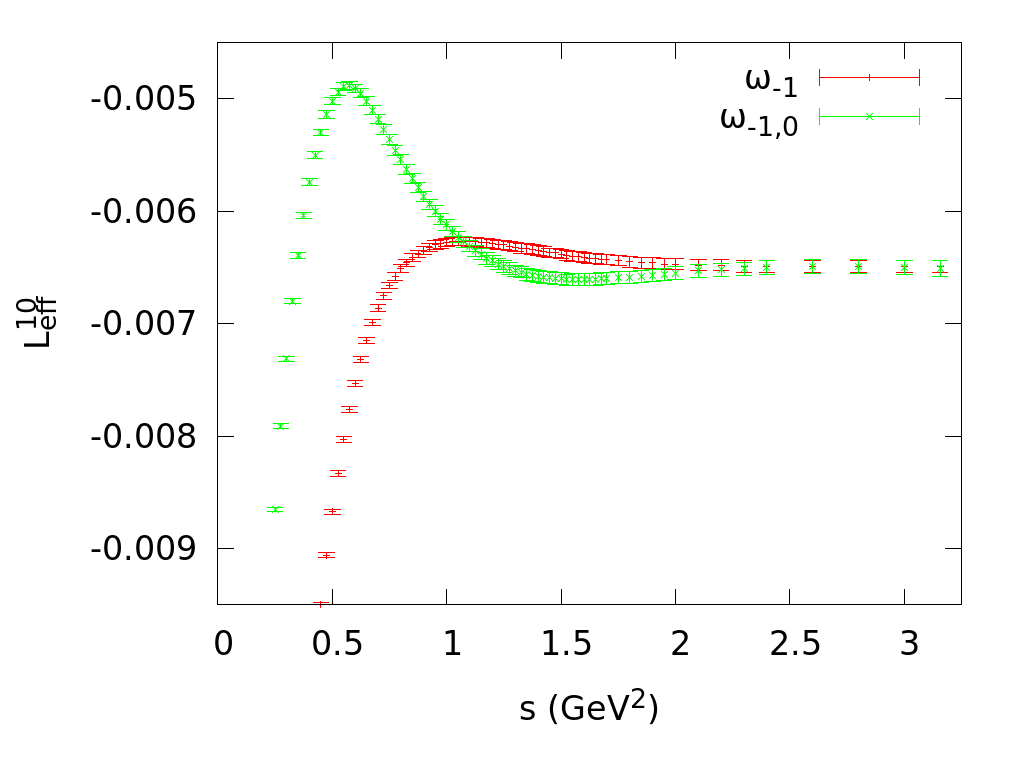}
\hfill
\includegraphics[width=0.48\textwidth]{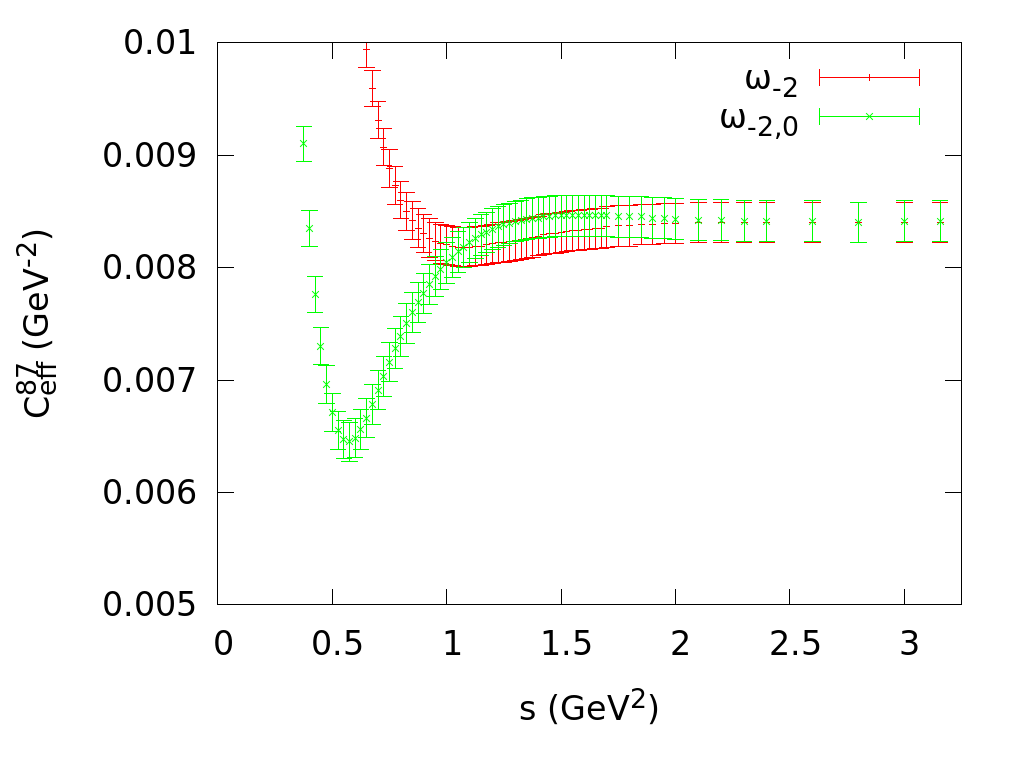}
\caption{$\Leff$ and $\Ceff$ at different values of $s_0$, using pinched weight functions and neglecting DVs.}
\label{fig.c87l102}
\end{figure}

Instead of weights of the form $s^{n}$, we can try to reduce DV effects using pinched weight functions \cite{LeDiberder:1992fr,Cirigliano:2003kc,Dominguez:2016jsq}, which vanish at $s=s_{0}$ (or in the vicinity) where the OPE breaks down. Following Ref.~\cite{GonzalezAlonso:2008rf}, we employ the WSRs in Eqs.~\eqn{wsr1} and \eqn{wsr2} and take $\omega_{-1,0}(s)=s^{-1} (1-s/s_0)$
and $\omega_{-1}(s)=s^{-1} (1-s/s_0)^2$
for estimating $\Leff$, and
$\omega_{-2,0}(s)=s^{-2} (1-s^2/s_0^2)$
and $\omega_{-2}(s)=s^{-2} (1-s/s_0)^2 (1+2 s/s_0)$
for estimating $\Ceff$.
Again, neglecting the DV terms, we plot the values of the effective couplings for different $s_{0}$ in Fig.~\ref{fig.c87l102}. We observe that using these pinched weights the results converge and become stable below $s=m_{\tau}^{2}$. This suggests that DV effects are negligible at $s_{0} \sim m_{\tau}^{2}$, when these pinched weight functions are used. Assuming that, we obtain:
\beqn\label{neg1}
\Leff & =& -(6.49\pm 0.05)\cdot 10^{-3} \, ,
\\[5pt]
\Ceff& =& (8.40\pm 0.18)\cdot 10^{-3}\; \mathrm{GeV}^{-2}\, .
\label{neg2}
\eeqn

\section{Dealing with violations of quark-hadron duality}
\label{dualsection}

The stability under changes of $s_0$ of the $\Leff$ and $\Ceff$ determinations is a necessary condition for vanishing duality violations. However the plateau could be accidental and disappear at slightly higher values of $s_0$ where experimental data are not available. Although this possibility looks rather unlikely, we want to gain confidence on our numerical results and perform a reliable estimation of the uncertainties associated with violations of duality, using Eq.~\eqn{eq:DV}. The problem is that the spectral function is experimentally unknown above $s=m_{\tau}^{2}$.

Fortunately, there are strong theoretical constraints on $\rho(s)$ that originate in the
special chiral-symmetry-breaking properties of $\Pi(s)$, implying its very fast fall-off  at large momenta. In addition to the two WSRs, the spectral function should satisfy the so-called Pion Sum Rule ($\pi$SR), which determines the electromagnetic pion mass splitting in the chiral limit \cite{Das:1967it}:
\bel{eq:PSR}
\int^{\infty}_{s_{\mathrm{th}}}ds\; s\;\log{\left(\frac{s}{\Lambda^2}\right)}\;\left.\frac{1}{\pi}\operatorname{Im} \Pi(s)\right|_{m_q=0}
\; =\; \left(m_{\pi^0}^2-m_{\pi^+}^2\right)_{\mathrm{em}}\; \frac{8\pi}{3\alpha}\; \left. f_\pi^2\right|_{m_q=0} \, .
\ee
Owing to the second WSR, the $\pi$SR does not depend on the arbitrary scale $\Lambda$. The r.h.s of this equation is well-known in $\chi$PT and, within the needed accuracy, we can identify in the l.h.s the spectral function in the chiral limit with the physical $\rho(s)$ because $m_q$ corrections are tiny.

\subsection{Parametrization of the spectral function}

All the theoretical and phenomenological knowledge we have about $\Pi(s)$ can be used to get an estimate of the DV uncertainties. In order to do that, let us adopt the following ansatz for the spectral function at large values of $s$ \cite{Cata:2008ru,GonzalezAlonso:2010rn,GonzalezAlonso:2010xf}:
\begin{equation}\label{eqparam}
\rho(s>s_{z})\; =\;\kappa\; e^{-\gamma s}\, \sin{\!\left\{\beta (s-s_{z})\right\}}\, ,
\end{equation}
with four free parameters $\kappa$, $\gamma$, $\beta$ and $s_z$.
This parametrization incorporates the expected strong fall-off when $s\rightarrow \infty $ and the oscillating behaviour predicted in resonance-based models \cite{Shifman:2000jv,Blok:1997hs,Shifman:1998rb}. We will split the spectral integrations in two parts, using the experimental data in the lower energy range and the ansatz~\eqn{eqparam} at higher energies. From the ALEPH data we know that the $V -A$
spectral function has a zero around $s_{z}\sim 2~\mathrm{GeV}^2$, which is represented
in Eq.~\eqn{eqparam} through the $s_z$ parameter. We will take this zero as the separation
point between the use of the data and the use of the model.

Our parametrization is compatible with the ALEPH spectral function above $s_z$. Fitting the parameters given in (\ref{eqparam}) to the ALEPH data in the interval $s\in(1.7 $ GeV$^{2},m_{\tau}^{2})$, we obtain a very good fit with $\chi^{2}_{\mathrm{min}}/\mathrm{d.o.f.} = 8.52/9$.
In fact, the fit with the updated ALEPH data looks more reliable compared to the previous one, where a value of $\chi^{2}_{\mathrm{min}}/\mathrm{d.o.f.}\ll 1$ was obtained \cite{GonzalezAlonso:2010rn}.

We want to stress that the exact $s$-dependence of the spectral function in the high-energy region cannot be derived from first principles. The ansatz~\eqn{eqparam} is just a convenient parametrization, consistent with present knowledge, that we are going to use to estimate theoretical uncertainties associated with violations of quark-hadron duality. Imposing that $\rho(s)$ should satisfy all known theoretical and experimental constraints, the free parameters in the ansatz will allow us to measure how much freedom remains for the spectral function shape and, therefore, to obtain a reliable estimate of the associated uncertainty.

There is an inherent systematic error in any work that estimates DV effects, namely the dependence on the chosen parametrization. The comparison with other works that parametrize the data in a different way, such as Refs.~\cite{Boito:2015fra,Boito:2012nt}, represents an important step in this regard.

\subsection{Selection of acceptable spectral functions}

Following the procedure described in \cite{GonzalezAlonso:2010rn}, we generate $3\cdot 10^{6}$ tuples of the parameters $(\kappa,\gamma,\beta,s_{z})$, randomly distributed in a rectangular region large enough to contain all the possible acceptable tuples. Among all generated tuples, we select those satisfying the following four physical conditions:

\begin{itemize}
\item[--] The tuples must be consistent with the ALEPH data above $s= 1.7~\mathrm{GeV}^2$, {\it i.e.}, they must be contained within the 90\% C.L. region in the fit to the experimental ALEPH spectral function described before:
\begin{equation}
\chi^{2}\; <\;\chi^{2}_{\mathrm{min}}\, +\, 7.78\; =\; 16.30\, .
\end{equation}
Although we will only use the ansatz above $s_{z}\sim 2~\mathrm{GeV}^2$,
we impose the compatibility with the data from $1.7~\mathrm{GeV}^2$ to ensure the continuity of the spectral function in the matching region between the data and the model.

\item[--] The tuples must satisfy within the experimental uncertainties up to $s_{z}$ the first and second WSRs with:
\beqn
\int^{s_{z}}_{0} ds\;\rho(s)^{\mathrm{ALEPH}}\;  +\; \int^{\infty}_{s_{z}} ds\;
\rho(s;\kappa,\gamma,\beta,s_{z})& =& 17.0\cdot 10^{-3}\,\text{GeV}^{2}\, ,\quad
\\[5pt]
\int^{s_{z}}_{0} ds\; s\;\rho(s)^{\mathrm{ALEPH}}\, +\, \int^{\infty}_{s_{z}} ds\; s\; \rho(s;\kappa,\gamma,\beta,s_{z}) & = & 0.24 \cdot 10^{-3}\,\text{GeV}^{4}\, ,
\eeqn
where the right-hand-side errors are omitted as they are negligible compared to the left-hand-side ones.

\item[--] The tuples must satisfy within the experimental uncertainties the $\pi$SR:
\beqn
\int^{s_{z}}_{0} ds\; s\;\log{\left(\frac{s}{1 \text{GeV}^{2}}\right)}\; \rho(s)^{\mathrm{ALEPH}} &+& \int^{\infty}_{s_{z}}
ds\; s\;\log{\left(\frac{s}{1 \text{GeV}^{2}}\right)}\;
\rho(s;\kappa,\gamma,\beta,s_{z})
\no\\[5pt]
 & = & - (10.9\pm 1.3) \cdot 10^{-3}\, \text{GeV}^{4}\, .
\eeqn
The quoted error in the $\pi$SR takes into account that quark masses do not vanish in nature and we are using real data instead of chiral-limit one. We estimate this uncertainty taking for the pion decay constant the range $f_{0} = (87 \pm 5)~\mathrm{MeV}$ \cite{GonzalezAlonso:2010rn}, which includes the physical value and its estimated value in the chiral limit \cite{Aoki:2013ldr}. We also include a small uncertainty coming from the
residual scale dependence of the logarithm, which is proportional to the second WSR.
\end{itemize}

\begin{figure}[tbh]\centering
\includegraphics[width=0.65\textwidth]{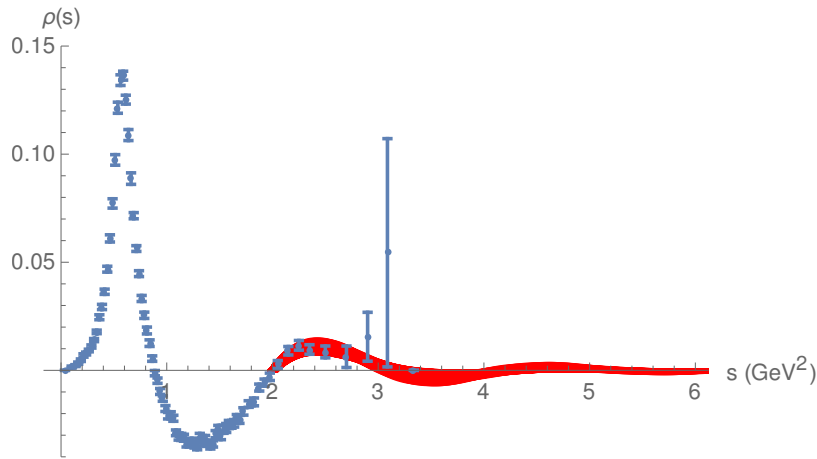}
\caption{Updated ALEPH $V-A$ spectral function~\cite{Davier:2013sfa} (blue points) and all the \lq\lq acceptable\rq\rq\ spectral functions (red band above 1.7~$\mathrm{GeV}^2$) that follow our parametrization and satisfy the physical conditions described in the main text.}
\label{fig:allfunctions}\end{figure}

\begin{figure}[tb]
\begin{center}
\includegraphics[width=0.48\textwidth]{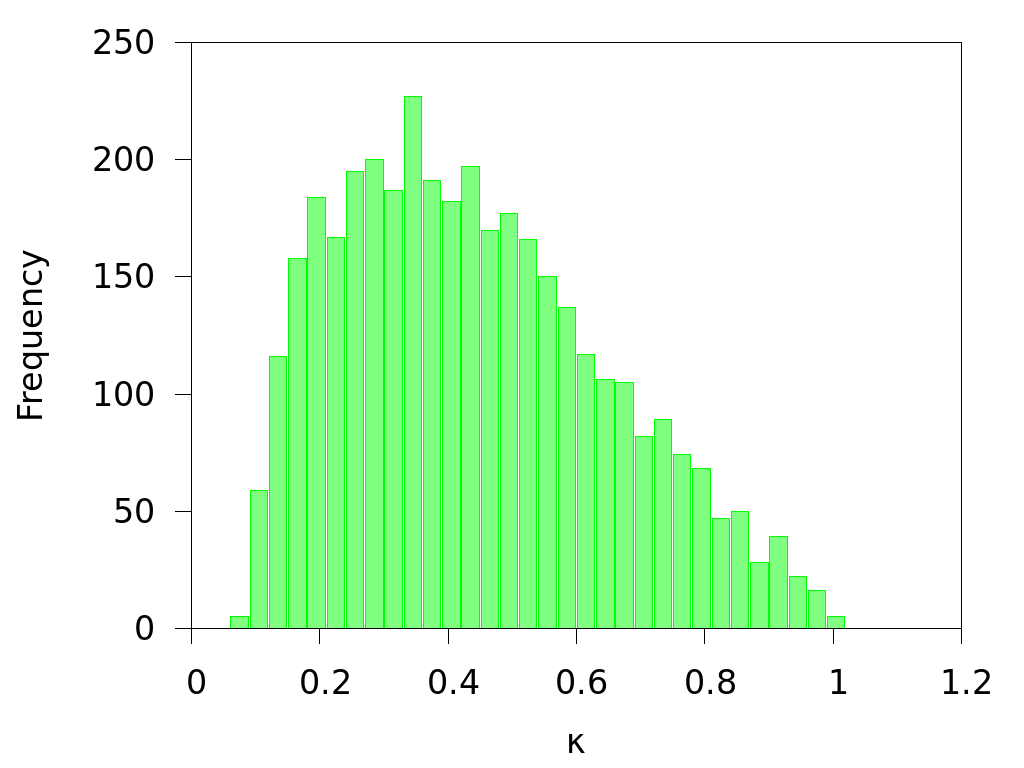}
\includegraphics[width=0.48\textwidth]{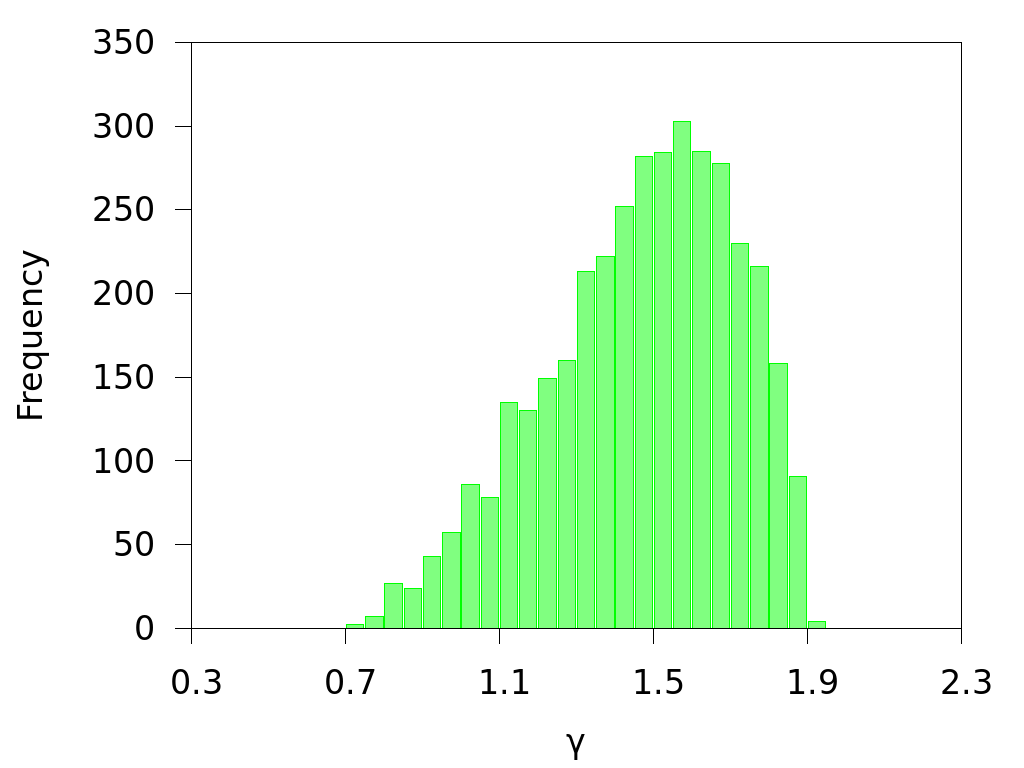}\\
\includegraphics[width=0.48\textwidth]{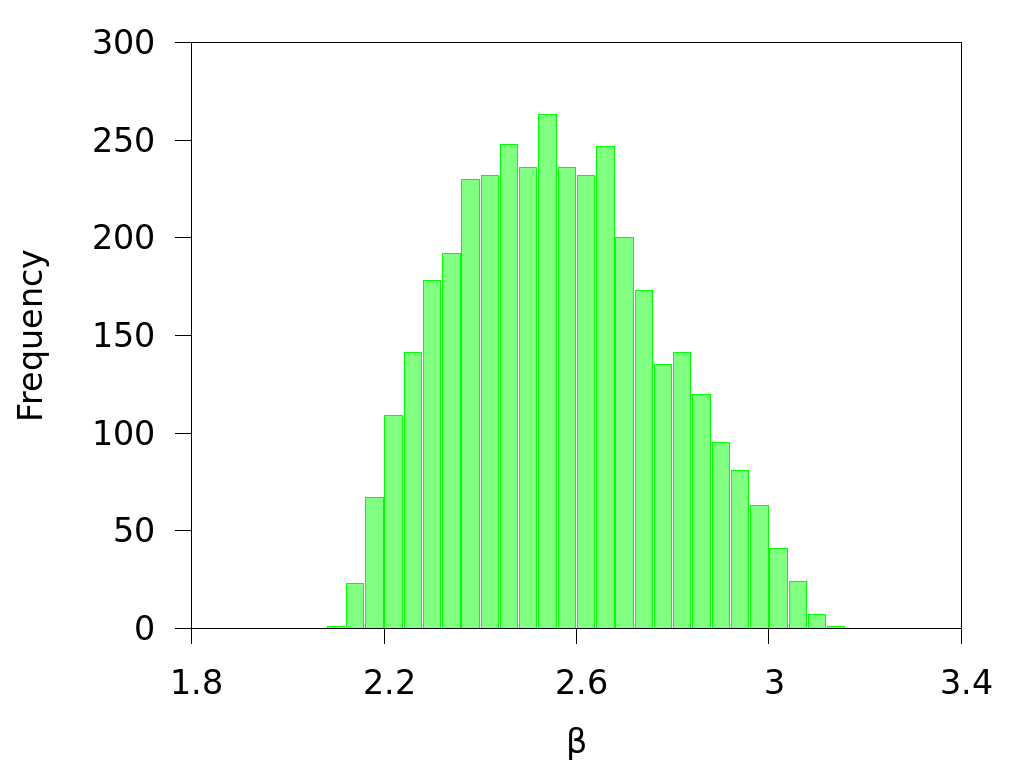}
\includegraphics[width=0.48\textwidth]{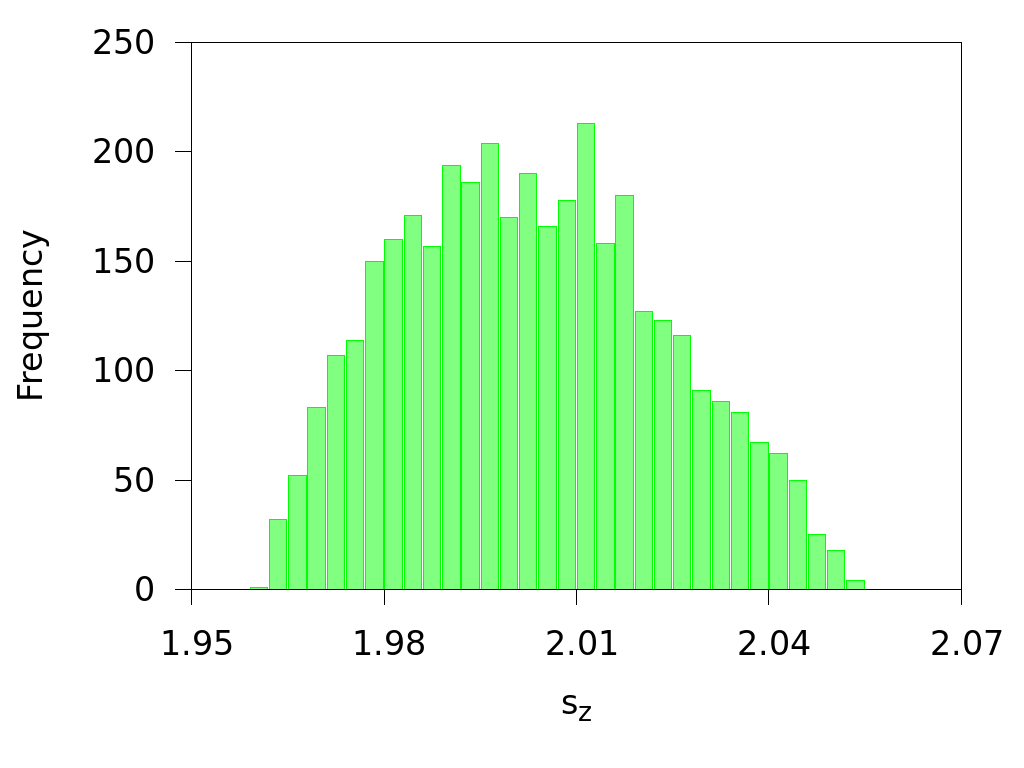}
\end{center}
\vskip -.5cm
\caption{Distributions of the parameters ($\kappa$, $\gamma$, $\beta$, $s_{z}$) that satisfy the physical constraints. GeV units are used for dimensionful quantities.}
\label{figparam}
\end{figure}

We accept only those tuples that fulfil the four conditions. This requirement constrains the regions in the parameter space of the ansatz~\eqn{eqparam} that are compatible with both QCD and the data. From the initial set of $3\cdot 10^{6}$ randomly generated tuples we obtain 3716 satisfying our set of minimal conditions. They represent the possible shapes of the spectral function beyond $s_z$, as shown in Fig.~\ref{fig:allfunctions}. In Fig.~\ref{figparam}, we plot the statistical distribution of the parameters $(\kappa,\gamma,\beta,s_{z})$ for the accepted tuples.

\section{Determination of physical parameters, including DV uncertainties}
\begin{figure}[t]\begin{center}
\includegraphics[width=0.48\textwidth]{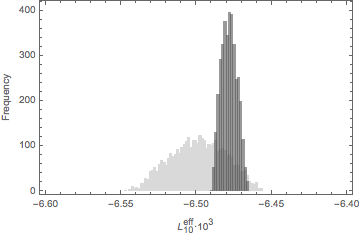}
\includegraphics[width=0.48\textwidth]{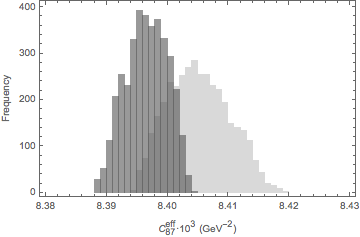}
\\[12pt]
\includegraphics[width=0.48\textwidth]{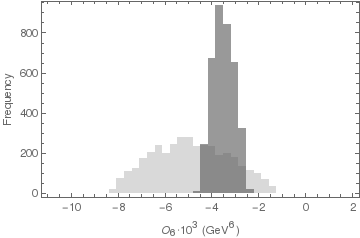}
\includegraphics[width=0.48\textwidth]{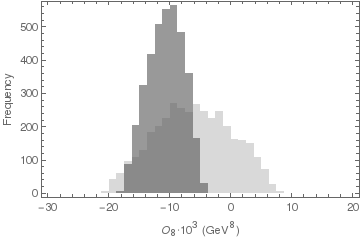}
\end{center}
\vskip -.5cm
\caption{Statistical distribution of $\Leff$, $\Ceff$, $\mathcal{O}_{6}$ y $\mathcal{O}_{8}$ for the tuples accepted, using $s^n$ weights (light gray) and pinched weight (dark gray) functions.}
\label{dist1}
\end{figure}

For every selected tuple we have an acceptable spectral function\footnote{Given by the ALEPH data below $s_{z}$ and by the parametrization~\eqn{eqparam} above that value.} that can be used
to estimate the different physical parameters through the corresponding spectral integrals. Using Eqs.~(\ref{eql10}), (\ref{eqc87}) and \eqn{eq:On} (for $n=2,3$) with $s_{0}=s_{z}$,
we determine $\Leff$, $\Ceff$, $\cO_6$ and $\cO_8$ for each of the 3716 accepted tuples. The statistical distributions of the calculated parameters are shown in Fig.~\ref{dist1} (light gray).

We can reduce both the experimental and the DV uncertainties using the following pinched weight functions \cite{GonzalezAlonso:2010xf}:
\beqn
\int^{s_{0}}_{s_{\mathrm{th}}}ds\;\frac{\rho(s)}{s^{2}}
\left(1-\frac{s}{s_{0}}\right)^{2} \left(1+\frac{2s}{s_{0}}\right) &\! = &\!
16\, \Ceff -6\,\frac{f_{\pi}^{2}}{s_{0}^{2}}
+4\,\frac{f_{\pi}^{2}m_{\pi}^{2}}{s_0^{3}}
-\delta_{\mathrm{DV}}[\omega_{-2},s_{0}]\, ,
\\[5pt]
\int^{s_{0}}_{s_{\mathrm{th}}}ds\;\frac{\rho(s)}{s}\left(1-\frac{s}{s_{0}}\right)^{2} &\! = &\!
-8\, \Leff-4\,\frac{f_{\pi}^{2}}{s_{0}}
+2\,\frac{f_{\pi}^{2}m_{\pi}^{2}}{s_{0}^{2}}-\delta_{\mathrm{DV}}[\omega_{-1},s_{0}]\, ,
\\[5pt]
\int^{s_{0}}_{s_{\mathrm{th}}}ds\;\rho(s)\left(s-s_{0}\right)^{2} &\! = &\!
2 f_{\pi}^{2} s_{0}^{2}-4f_{\pi}^{2}m_{\pi}^{2}\,s_{0}+2 f_{\pi}^{2}m_{\pi}^{4}+\mathcal{O}_{6}\label{o6eq}
-\delta_{\mathrm{DV}}[\omega_{2},s_{0}]\, ,
\\[5pt]
\int^{s_{0}}_{s_{\mathrm{th}}}ds\;\rho(s)\left(s-s_{0}\right)^{2}(s+2s_{0}) &\! = &\!
-6 f_{\pi}^{2}m_{\pi}^{2}s_{0}^{2}+4f_{\pi}^{2}s_{0}^{3}+2 f_{\pi}^{2}m_{\pi}^{6}-\mathcal{O}_{8}\label{o8eq}
-\delta_{\mathrm{DV}}[\omega_{3},s_{0}]\, .\qquad
\eeqn
Following the same method with these relations, we obtain new distributions of acceptable physical parameters, which are also shown in Fig.~\ref{dist1} (dark gray). From these new distributions we get:
\beqn
\Leff&= & (-6.477 \,{}^{+\, 0.004}_{-\, 0.006} \pm 0.05) \cdot 10^{-3}
\; =\; (-6.48 \pm 0.05) \cdot 10^{-3}\, ,
\label{resultpw1}\\[5pt]
\Ceff&=& (8.399\,{}^{+\, 0.002}_{-\, 0.005} \pm 0.18) \cdot 10^{-3}\, \mathrm{GeV}^{-2}
\; =\; (8.40  \pm 0.18) \cdot 10^{-3}\; \mathrm{GeV}^{-2}\, ,
\label{resultpw1.5}\\[5pt]
\mathcal{O}_{6}&=& (-3.6 \,{}^{+\, 0.5}_{-\, 0.4} \pm 0.5) \cdot 10^{-3}\; \mathrm{GeV}^{6}
\; =\; (-3.6 \,{}^{+\, 0.7}_{-\, 0.6}) \cdot 10^{-3}\; \mathrm{GeV}^{6}\, ,
\label{resultpw1.9}\\[5pt]
\mathcal{O}_{8}&=& (-1.0 \pm 0.3 \pm 0.2) \cdot 10^{-2}\; \mathrm{GeV}^{8}
\; =\; (-1.0 \pm 0.4) \cdot 10^{-2}\; \mathrm{GeV}^{8}\, ,
\label{resultpw2}
\eeqn
where the first errors correspond to DV uncertainties, computed from the dispersion of the histograms, and the second errors are the experimental ones.

We observe that pinched weight functions reduce indeed the DV effects, and that they are negligible for $\Leff$ and $\Ceff$ at $s_{0}\sim m_{\tau}^{2}$, compared with the experimental uncertainties. The results obtained for these two LECs are in perfect agreement with our first determinations
in Eqs.~\eqn{neg1} and \eqn{neg2} that did not include any estimate of DV. The corresponding spectral integrals contain weight functions with negative powers of $s$ that suppress the contribution from the upper end of the integration range, making DV irrelevant. This is no-longer true for the vacuum condensates $\cO_6$ and $\cO_8$, which are determined with weight functions growing with positive powers of $s$. The use of pinched weights is then essential to suppress the contributions from the region around $s_0$ in the contour integration. This is clearly reflected
in the strong reduction of uncertainties observed in the two lower panels of Fig.~\ref{dist1}.

\begin{figure}[t]\begin{center}
\includegraphics[width=0.48\textwidth]{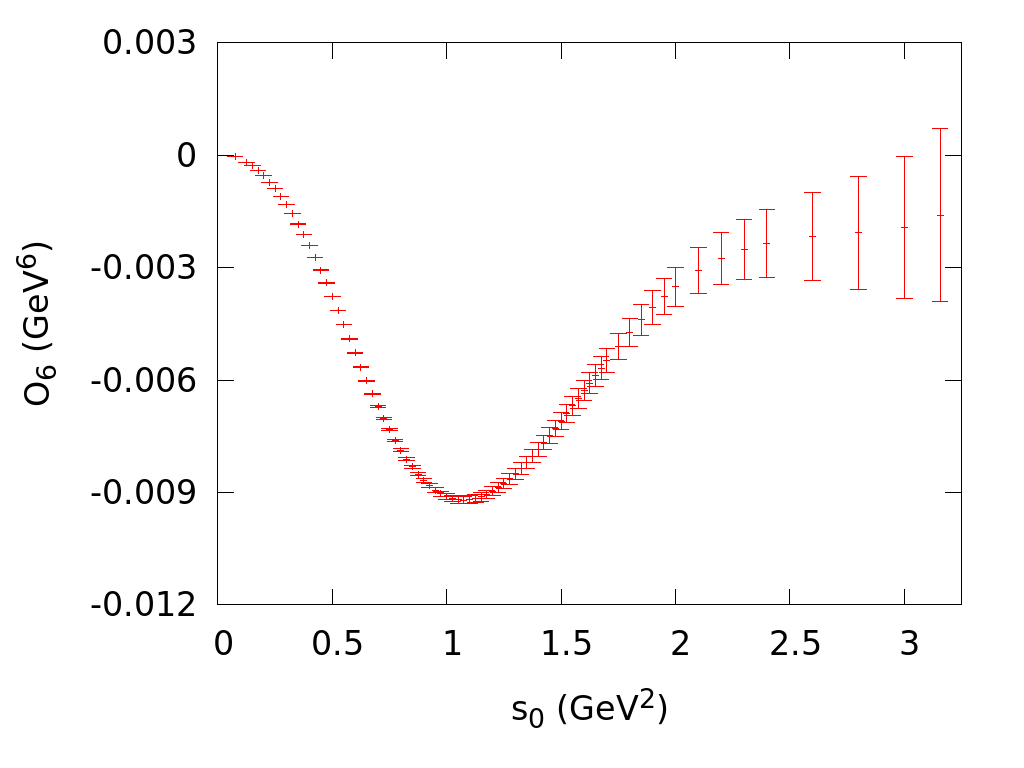}
\includegraphics[width=0.48\textwidth]{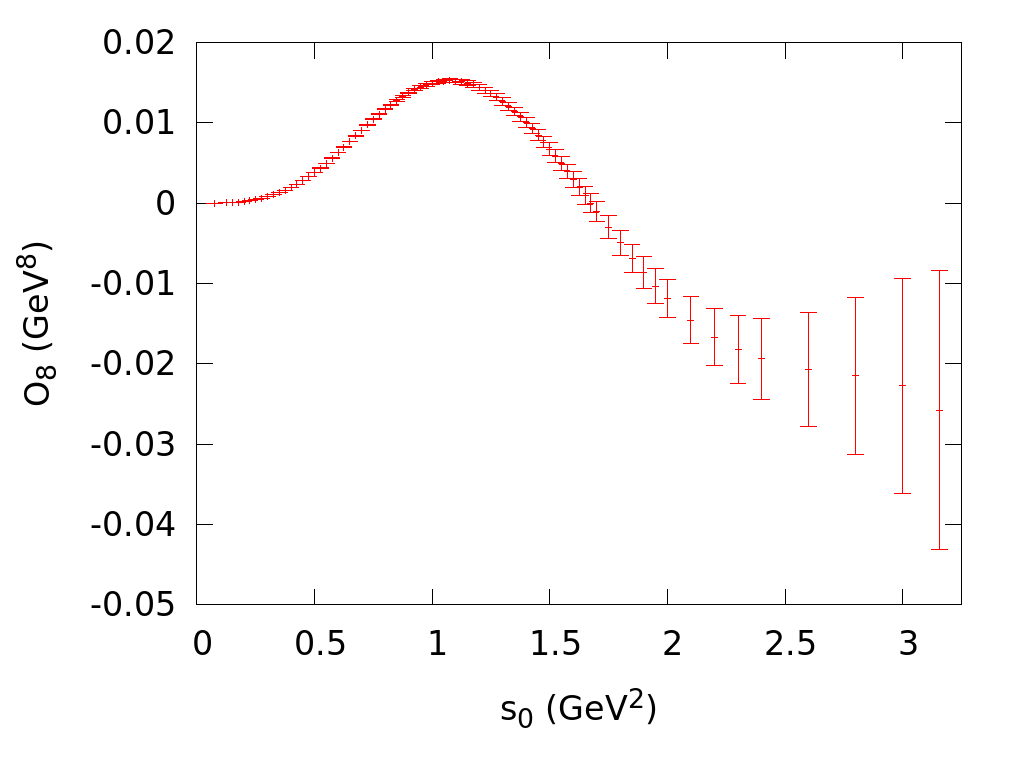}
\end{center}
\vskip -.5cm
\caption{Values of the condensates $\mathcal{O}_{6}$ and $\mathcal{O}_{8}$, at different values of $s_0$, obtained from Eqs.~$\eqn{o6eq}$ and $\eqn{o8eq}$ ignoring duality violations.}
\label{o68}
\end{figure}

Actually, ignoring completely DV effects, from the double-pinched weight functions in Eqs.~$\eqn{o6eq}$ and $\eqn{o8eq}$ one obtains values for $\cO_6$ and $\cO_8$ that are perfectly compatible with our determinations in Eqs.~\eqn{resultpw1.9} and \eqn{resultpw2}, although with much larger experimental uncertainties. This is illustrated in Fig.~\ref{o68}, which shows how the extracted condensates stabilize at large $s_0$, around the right values but with very large error bars. The implementation of short-distance constraints (WSRs and $\pi$SR), through the procedure described in the previous section, has made possible to better pin down the spectral function in that region and obtain the more precise values in Eqs.~\eqn{resultpw1.9} and \eqn{resultpw2}.

Our results are in good agreement with those obtained previously in Ref.~\cite{GonzalezAlonso:2010xf} with the 2005 ALEPH data set. Thus, the improvements incorporated in the 2014 release of the ALEPH data do not introduce sizeable modifications of the physical outputs. Similar results have been obtained recently in Ref.~\cite{Boito:2015fra}, using also the 2014 ALEPH data set.

Ref.~\cite{Boito:2015fra} emphasises the existence of a slight tension with the results obtained in Ref.~\cite{Boito:2012nt} with the 1999 OPAL data set \cite{Ackerstaff:1998yj}.
In view of this, we have repeated our numerical analyses with the OPAL spectral
function \cite{Ackerstaff:1998yj}. As happened with the 2005 ALEPH data set,
the fit of the ansatz (\ref{eqparam}) to the OPAL data in the interval $s \in (1.7\;\mathrm{GeV}^2, m_\tau^2)$ has a $\chi_{\mathrm{min}}/\mathrm{d.o.f.}\ll 1$. Applying the same procedure used for ALEPH, we have obtained the following results with the OPAL data:
\beqn
\label{resultpwopal}
\Leff &= & (-6.42 \pm 0.10) \cdot 10^{-3}\, ,
\\[5pt]
\Ceff &=& (8.35 \pm 0.29) \cdot 10^{-3}\; \mathrm{GeV}^{-2}\, ,
\\[5pt]
\mathcal{O}_{6} &=& (-5.7 ^{+1.1}_{-1.2}) \cdot 10^{-3}\; \mathrm{GeV}^{6}\, ,
\\[5pt]
\mathcal{O}_{8} &=& (0.0 ^{+0.5}_{-0.6}) \cdot 10^{-2}\; \mathrm{GeV}^{8}\, .
\label{resultpwopal2}
\eeqn
Owing to the larger uncertainties of the OPAL data, specially at higher values of $s$, the extracted parameters are less precise than those obtained with the ALEPH data. Nevertheless, comparing Eqs. (\ref{resultpwopal})-(\ref{resultpwopal2}) with (\ref{resultpw1})-(\ref{resultpw2}), we observe a good agreement between both sets of results, the differences being only $0.5\,\sigma$, $0.1\,\sigma$, $1.6\,\sigma$ and $1.4\,\sigma$ for
$\Leff$, $\Ceff$, $\mathcal{O}_{6}$ and $\mathcal{O}_{8}$, respectively.
We conclude that the much larger fluctuations obtained in Refs.~\cite{Boito:2015fra,Boito:2012nt} between the results extracted from the two data sets are a consequence of the particular approach adopted in their DV analyses, which does not look optimal to us.\footnote{
In Refs.~\cite{Boito:2012nt} and \cite{Boito:2015fra} the exact $s$-dependence of the resonance-based model (\ref{eqparam}) is assumed to be true for the $V$ and $A$ channels separately and a complex analysis involving 9 parameters, including a model-dependent determination of the strong coupling, is performed. In this way, uncertainties related to an $\alpha_s$ determination from the $V$ and $A$ spectral distributions are introduced in the analysis of the correlator $\Pi(s)$, which does not contain any perturbative contribution. Moreover the LECs and vacuum condensates are directly extracted from the fitted $V$ and $A$ spectral functions without imposing any further requirement
(WSRs and $\pi$SR are only checked to be satisfied within errors a posteriori).
Since DV is not very relevant for the extraction of the LECs, similar values are obtained for $\Leff$ and $\Ceff$ with the two data sets. However, sizeable differences show up in their determinations of $\cO_6$ and $\cO_8$ where DV is more important.}

Finally, we can use double-pinched weight functions in order to estimate higher-dimensional condensates:
\beqn
\lefteqn{\int^{s_{0}}_{s_{\mathrm{th}}}ds\;\rho(s)\;\left(s-s_{0}\right)^{2}(s^{2}+2s_{0}s+3s_{0}^{2})}
&&\no\\
&\hskip 3cm =&\mbox{} -8f_{\pi}^{2}m_{\pi}^{2}s_{0}^{3}+6 f_{\pi}^{2}s_{0}^{4}+2f_{\pi}^{2}m_{\pi}^{8}+\mathcal{O}_{10}
-\delta_{\mathrm{DV}}[\omega_{4},s_{0}]\, ,
\\[7pt]
\lefteqn{\int^{s_{0}}_{s_{\mathrm{th}}}ds\;\rho(s)\;\left(s-s_{0}\right)^{2}(s^{3}+2s_{0}s^{2}+3s_{0}^{2}s+4s_{0}^{3})}
&&\no\\
&\hskip 3cm =&\mbox{} -10f_{\pi}^{2}m_{\pi}^{2}s_{0}^{4}+8 f_{\pi}^{2}s_{0}^{5}+2f_{\pi}^{2}m_{\pi}^{10}-\mathcal{O}_{12}
-\delta_{\mathrm{DV}}[\omega_{5},s_{0}]\, ,
\\[7pt]
\lefteqn{\int^{s_{0}}_{s_{\mathrm{th}}}ds\;\rho(s)\;\left(s-s_{0}\right)^{2}(s^{4}+2s_{0}s^{3}+3s_{0}^{2}s^{2}+4s_{0}^{3}s+5s_{0}^{4})}
&&\no\\
&\hskip 3cm =&\mbox{} -12f_{\pi}^{2}m_{\pi}^{2}s_{0}^{5}+10 f_{\pi}^{2}s_{0}^{6}+2f_{\pi}^{2}m_{\pi}^{12}+\mathcal{O}_{14}
-\delta_{\mathrm{DV}}[\omega_{6},s_{0}]\, ,
\\[7pt]
\lefteqn{\int^{s_{0}}_{s_{\mathrm{th}}}ds\;\rho(s)\;\left(s-s_{0}\right)^{2}(s^{5}+2s_{0}s^{4}+3s_{0}^{2}s^{3}+4s_{0}^{3}s^{2}+5s_{0}^{4}s+6s_{0}^{5})}
&&\no\\
&\hskip 3cm =&\mbox{} -14f_{\pi}^{2}m_{\pi}^{2}s_{0}^{6}+12 f_{\pi}^{2}s_{0}^{7}+2f_{\pi}^{2}m_{\pi}^{14}-\mathcal{O}_{16}
-\delta_{\mathrm{DV}}[\omega_{7},s_{0}]\, .
\eeqn
Using these equations with the same method, we obtain from the ALEPH data:
\beqn\label{eqhigher1}
\mathcal{O}_{10}&=& (5.6 \pm 1.2 \pm 0.8) \cdot 10^{-2}\; \mathrm{GeV}^{10}\; =\;
      (5.6 \pm 1.4) \cdot 10^{-2}\; \mathrm{GeV}^{10}\, ,
\\[5pt]
\mathcal{O}_{12}&=& (-0.13\,{}^{+\, 0.01}_{-\, 0.06} \pm 0.02) \; \mathrm{GeV}^{12}
\; =\; (-0.13\,{}^{+\, 0.02}_{-\, 0.07}) \; \mathrm{GeV}^{12}\, ,
\\[5pt]
\mathcal{O}_{14}&=& (0.24\,{}^{+\, 0.11}_{-\, 0.05} \pm 0.06)\; \mathrm{GeV}^{14}\;
=\; (0.24 \,{}^{+\, 0.12}_{-\, 0.08}) \; \mathrm{GeV}^{14}\, ,
\\[5pt]
\mathcal{O}_{16}&=& (-0.38\,{}^{+\, 0.25}_{-\, 0.10} \pm 0.13) \; \mathrm{GeV}^{14}\;
=\; (-0.38\,{}^{+\, 0.28}_{-\, 0.17}) \; \mathrm{GeV}^{16}\, .
\label{eqhigher4}
\eeqn

\subsection{Comparison with previous works}

Our final results for $\Leff$,  $\Ceff$, $\cO_{6}$ and $\cO_{8}$ are compared in Table~\ref{tab:compilation} with recent (post-2005) phenomenological determinations of these parameters, obtained with different data sets~\cite{Ackerstaff:1998yj,Schael:2005am,Davier:2013sfa} and various DV parametrizations.\footnote{A complete list including theoretical estimates \cite{Friot:2004ba,Masjuan:2007ay,Pich:2008jm} and previous phenomenological determinations of these quantities (and of higher-dimensional condensates) \cite{Cirigliano:2002jy,Cirigliano:2003kc,Schael:2005am,Dominguez:2003dr,Ackerstaff:1998yj,Barate:1998uf,Davier:1998dz,Bijnens:2001ps,Ioffe:2000ns,Geshkenbein:2001mn,Zyablyuk:2004iu,Rojo:2004iq,Narison:2004vz,Peris:2000tw,Almasy:2006mu} can be found in Refs.~\cite{GonzalezAlonso:2010xf,MGAthesis}.}

\begin{table}[tb]
\begin{center}
\renewcommand\arraystretch{1.2}
\begin{tabular}{|c|c|c|c|c|c|}
\hline
$10^3\cdot \Leff$ & $10^3\cdot \Ceff$ &
$10^3\cdot \cO_{6}$ & $10^2\cdot \cO_{8}$ &
Reference & Comments
\\
& {\small $(\mathrm{GeV}^{-2})$} & {\small $(\mathrm{GeV}^6)$} & {\small $(\mathrm{GeV}^8)$} & &
\\
\hline
$-6.45\pm 0.06$ & -- & $-2.3\pm 0.6$ & $-5.4\pm 3.3$ &
BPDS'06 \protect\cite{Bordes:2005wv} & ALEPH'05\,+\,DV$=\!0$
\\
-- & -- & $-6.8 \, {}^{+\, 2.0}_{-\, 0.8}$ & $3.2 \, {}^{+\, 2.8}_{-\, 9.2}$ &
ASS'08 \protect\cite{Almasy:2008xu} & ALEPH'05\,+\,DV$=\!0$
\\
$-6.48\pm 0.06$ & $8.18\pm 0.14$ & -- & -- &
GPP'08 \protect\cite{GonzalezAlonso:2008rf} & ALEPH'05\,+\,DV$=\!0$
\\
$-6.44\pm 0.05$ & $8.17\pm 0.12$ & $-4.4\pm 0.8$ & $-0.7\pm 0.5$ &
GPP'10 \protect\cite{GonzalezAlonso:2010rn,GonzalezAlonso:2010xf} & ALEPH'05\,+\,DV$_{V\!-\!A}$
\\
$-6.45\pm 0.09$ & $8.47\pm 0.29$ & $-6.6\pm 1.1$ & $\phantom{-}0.5\pm 0.5$ &
Boito'12 
\protect\cite{Boito:2012nt} & OPAL'99\,+\,DV$_{V/A}$
\\
$-6.50\pm 0.10$ & -- & $-5.0\pm 0.7$ & $-0.9\pm 0.5$ &
DHSS'15 \protect\cite{Dominguez:2014fua} & ALEPH'14\,+\,DV$=\!0$
\\
$-6.45\pm 0.05$ & $8.38\pm 0.18$ & $-3.2\pm 0.9$ & $-1.3\pm 0.6$ &
Boito'15 \protect\cite{Boito:2015fra} & ALEPH'14\,+\,DV$_{V/A}$
\\ \hline
$-6.42\pm 0.10$ & $8.35\pm 0.29$ & $-5.7\, {}^{+\, 1.1}_{-\, 1.2}$ & $0.0\,{}^{+\, 0.5}_{-\, 0.6}$ &
this work & OPAL'99\,+\,DV$_{V\!-\!A}$
\\
$-6.48\pm 0.05$ & $8.40\pm 0.18$ & $-3.6\,{}^{+\, 0.7}_{-\, 0.6}$ & $-1.0 \pm 0.4$&
this work & ALEPH'14\,+\,DV$_{V\!-\!A}$
\\ \hline
\end{tabular}
\caption{Compilation of recent determinations of the LECs and vacuum condensates.}
\label{tab:compilation}
\end{center}
\end{table}

There is an excellent agreement among the different values quoted for the effective LECs $\Leff$ and $\Ceff$, showing that these determinations are very solid and do not get affected by DV effects. In fact, as shown in Table~\ref{tab:compilation}, the precision has not changed in the last ten years. Nonetheless, the robustness of these determinations has increased significantly thanks to the thorough studies of DV effects with different approaches. The values obtained from different data sets are also in good agreement, although one can notice a $1\,\sigma$ shift of the $\Ceff$ central value when changing from the old (2005) to the updated (2014) ALEPH data.

The different results for $\cO_{6}$ and $\cO_{8}$ are also in reasonable agreement, within the quoted uncertainties. A good control of DV effects is more important for these vacuum condensates. The use of pinched weights allows to sizeably reduce their impact and obtain more reliable determinations. With the ALEPH'14 data one reaches a 20\% accuracy for $\cO_{6}$, but the error remains still large (40\%) for $\cO_{8}$. As commented before, we do not see any significant discrepancy between the results obtained from the OPAL and ALEPH data samples.

\section{$\boldsymbol{\chi}$PT couplings}
\label{sec:ChPT}

The effective couplings $\Leff$ and $\Ceff$ can be rewritten in terms of $\cO(p^4)$ and $\cO(p^6)$ couplings of the $\chi$PT Lagrangian~\cite{GonzalezAlonso:2008rf,Amoros:1999dp}:
\beqn
\label{L10-p6}
\Leff &\!\equiv &\! -\frac{1}{8}\, \overline{\Pi}(0)
\nonumber\\ &\! = &\!
L_{10}^r(\mu) \, +\,
\frac{1}{128 \,\pi^2} \left[1- \log{\left(\frac{\mu^2}{m_\pi^2}\right)}\,
+ \,\frac{1}{3}\,\log{\left(\frac{m_K^2}{m_\pi^2}\right)} \right]
\nonumber\\ &\! -&\!
\frac{1}{8}\left( {\cal C}_0^r + {\cal C}_1^r \right)(\mu)
- 2\, \left( 2 \mu_\pi + \mu_K \right) \,\left( L_9^r + 2 L_{10}^r\right)\! (\mu)\,
+\,  G_{2L}(\mu,s\!=\!0) \, +\, {\cal O} (p^8) \, ,
\\[10pt]\label{C87-p6}
\Ceff &\!\equiv &\! \frac{1}{16}\, \overline{\Pi}\,{}'(0)
\nonumber\\ &\! = &\!
C_{87}^r(\mu) \, - \,
\frac{1}{64 \,\pi^2 f_\pi^2} \left[1- \log{\left(\frac{\mu^2}{m_\pi^2}\right)}\,
+ \,\frac{1}{3}\,\log{\left(\frac{m_K^2}{m_\pi^2}\right)} \right] \, L_9^r(\mu)
\nonumber\\ &\! + &\!
\frac{1}{7680\, \pi^2} \left( \frac{1}{m_K^2} + \frac{2}{m_\pi^2} \right)
\, -\frac{1}{2}\, G'_{2L}(\mu,s\!=\!0)  \, +\, {\cal O} (p^8)\, ,
\eeqn
where the factors $\mu_i= m_i^2 \log(m_i/\mu)/(16 \pi^2 f_\pi^2)$ originate from one-loop corrections and $G_{2L}(\mu,s\!=\!0)$ and $G'_{2L}(\mu,s\!=\!0)$ are two-loop functions, whose numerical values are given in the appendix. We have also defined
\beqn
{\cal C}_0^r &=& 32\, m_\pi^2\, (C_{12}-C_{61}+C_{80})\, ,
\\
{\cal C}_1^r &=& 32\, (m_\pi^2+2m_K^2)\, (C_{13}-C_{62}+C_{81})\, .
\eeqn
To first approximation the effective parameters correspond to the chiral couplings $L_{10}$ and $C_{87}$, which appear at $\cO(p^4)$ and $\cO(p^6)$, respectively, in the $\chi$PT expansion. The scale dependence of $L^r_{10}(\mu)$ is cancelled by the one-loop logarithmic terms in the second line of Eq.~\eqn{L10-p6}, which are suppressed by one power of $1/N_C$ with respect to $L_{10}^r(\mu)$, where $N_C$ is the number of QCD colours. The remaining contributions in Eq.~\eqn{L10-p6} contain the $\cO(p^6)$ corrections, which unfortunately introduce other $\cO(p^6)$ and $\cO(p^4)$ chiral couplings (third line).
The corrections to $C^r_{87}(\mu)$ in Eq.~\eqn{C87-p6} only involve one additional LEC, $L_9^r(\mu)$, through a one-loop correction with the $\cO(p^4)$ chiral Lagrangian.

It is convenient to give the following compact numerical form of these equations to ease their future use:
\beqn
\label{L10-p4-num}
\Leff &=& L_{10}^r -0.00126 \,+\, {\cal O} (p^6)~,\\
\label{L10-p6-num}
\Leff &=& 1.53\, L_{10}^r + 0.263\, L_9^r - 0.00179 - \frac{1}{8}\left( {\cal C}_0^r + {\cal C}_1^r \right)\! \, +\, {\cal O} (p^8)~,\\
\label{C87-p6-num}
\Ceff &=& C_{87}^r + 0.296\, L_9^r + 0.00155\, +\, {\cal O} (p^8)~,
\eeqn
where we have used $\mu=M_\rho$ as the reference value for the $\chi$PT renormalization scale. The uncertainties in these numbers are much smaller than those affecting the different LECs and can therefore be neglected.

Working with $\cO(p^4)$ precision, the determination of $L^r_{10}(\mu)$ is straightforward
and we find:
\be
\label{valL10p4}
L_{10}^r(M_\rho)\; =\; -(5.22 \pm 0.05) \cdot 10^{-3}  \hskip 1.5cm \mbox{[$\cO(p^4)$ analysis]}\, .
\ee

As mentioned before, an $\cO(p^6)$ determination of $L_{10}^r$ requires to know some next-to-next-to-leading-order (NNLO) LECs,\footnote{It also requires $L_9^r$, which we take from Ref.~\cite{Bijnens:2002hp}: $L_9^r (M_{\rho}) = 5.93\; (43) \cdot 10^{-3}$. Let us notice that this is the value used also in all other $\cO(p^6)$ extractions of $L_{10}^r$ from tau data.}
namely those in ${\cal C}_{0,1}^r$. This has motivated some interest in these quantities in the last few years. Here we briefly review the different approaches.

\begin{figure}[tb]\centering
\includegraphics[width=0.8\textwidth]{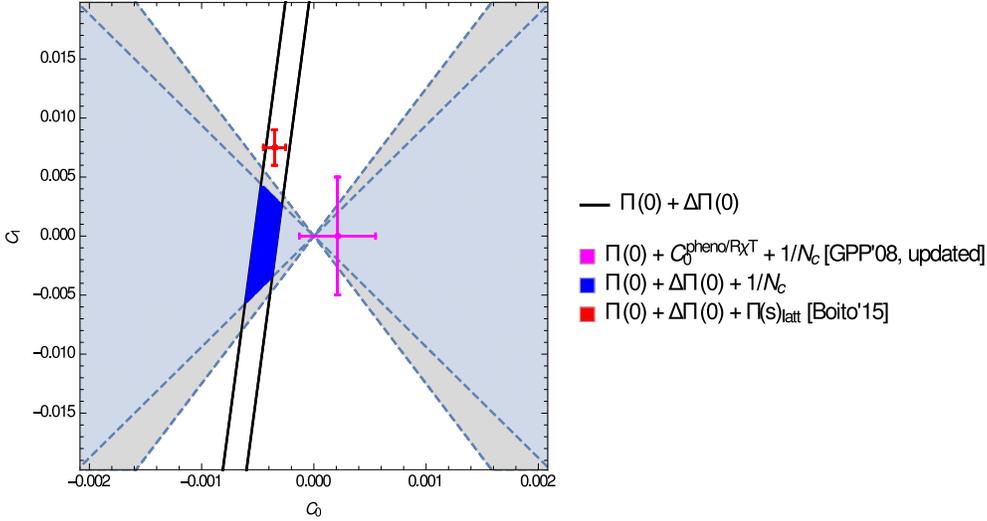}
\caption{Latest determinations of the linear combinations of NNLO LECs ${\cal C}_{0,1}^r$, at $\mu=M_\rho$. We follow the same notation as in Table~\ref{tab:compilation-bis}.
The region allowed by the inequality of Eq.~\eqref{Ncguess}, inspired by large-$N_c$ arguments, is indicated in light blue, whereas the light gray area around it (dashed) simply represents a naive estimate of its error, namely 33\%. }
\label{fig:C0C1plot}
\end{figure}

In the first $\cO(p^6)$ determination of $L_{10}^r$~\cite{GonzalezAlonso:2008rf}, ${\cal C}_0^r$ was extracted from a combination of phenomenological ($C_{61,12}^r$)~\cite{Durr:1999dp,Jamin:2004re,Kampf:2006bn,Golterman:2014nua} and theoretical ($C_{80}^r$, R$\chi$T)~\cite{Amoros:1999dp,Unterdorfer:2008zz} inputs, namely\footnote{
This value of $C_{61}^r$ comes from a flavour-breaking finite-energy sum rule involving the correlator $\overline\Pi^{(0+1)}_{ud-us,VV}(0)$. The original result~\cite{Durr:1999dp} has been updated recently~\cite{Golterman:2014nua}, finding
$$
32\, (m_K^2-m_\pi^2)\, C_{61} +1.06\, L_{10}^r\, =\, 0.00727\; (134)~.
$$
Since $L_{10}^r$ appears in this relation only at one loop, {\it i.e.} at $\cO(p^6)$, we can use here an $\cO(p^4)$ determination of $L_{10}^r$ to extract $C_{61}^r$. We can indeed see that the $L_{10}^r$ contribution to the $C_{61}^r$ error is subdominant. We use the conservative value $L_{10}^r=-0.0052\; (17)$ to extract $C_{61}^r$.}
%
\beqn
C_{61}^{r} (M_{\rho}) &=& (1.7 \pm 0.6) \cdot 10^{-3}\; \mathrm{GeV}^{-2}\quad \text{\cite{Kampf:2006bn,Durr:1999dp,Golterman:2014nua}}\, ,
\label{const1}\\
C_{12}^{r} (M_{\rho}) &=& (0.4 \pm 6.3) \cdot 10^{-5}\; \mathrm{GeV}^{-2}\quad  \text{\cite{Jamin:2004re}}\, ,
\label{const2}\\
C_{80}^{r} (M_{\rho}) &=& (2.1 \pm 0.5) \cdot 10^{-3}\; \mathrm{GeV}^{-2}\quad \text{\cite{Amoros:1999dp,Unterdorfer:2008zz}}\, ,
\label{const3}
\eeqn
whereas ${\cal C}_1^r$, which was completely unknown at the time, was estimated using
\beqn
|C_{62}^{r}-C_{13}^{r}-C_{81}^{r}|
\;\le\; \frac{1}{3}\, |C_{61}^{r}-C_{12}^{r}-C_{80}^{r}|\, ,
\label{Ncguess}
\eeqn
{\it i.e.}, a simple educated guess based on the fact that those LECs are suppressed by a factor $1/N_{C}$. 
Using these numbers and Eq. (\ref{L10-p6-num}), we obtain the results shown in Table~\ref{tab:compilation-bis} (5th row) and Fig.~\ref{fig:C0C1plot} (magenta point), which supersede those found in Ref.~\cite{GonzalezAlonso:2008rf}.

\begin{table}[tb]
\begin{center}
\renewcommand\arraystretch{1.2}
\begin{tabular}{|c|c|c|c|l|}
\hline
$L_{10}^r (M_\rho)$ & ${\cal C}_0^r(M_\rho)$ & ${\cal C}_1^r(M_\rho)$ & Reference & ~~~~Input
\\
$\times 10^3$		& $\times 10^3$	& $\times 10^3\,$	&  &
\\
\hline
-4.06 (39)	& +0.54 (42)	&	0 (5)	& GPP'08 \protect\cite{GonzalezAlonso:2008rf} & $\Pi(0)$ + ${\cal C}_0^{\mbox{\scriptsize{pheno/R$\chi$T}}}$ + $1/N_c$\\
-3.10 (80)	& -0.81 (82)	& 	14 (10)	& Boito'12 \protect\cite{Boito:2012nt} & $\Pi(0)$ + $\Pi(s)_{\rm{latt}}$\\
-3.46 (32)	& -0.34 (13)	& 	8.1 (3.5)& Boyle'14, GMP'14 \protect\cite{Golterman:2014nua,Boyle:2014pja} & $\Pi(0)$ + $\Pi(s)_{\rm{latt}}$ + $\Delta\Pi(0)$\\
-3.50 (17)	& -0.35 (10)	& 	7.5 (1.5)& Boito'15 \protect\cite{Boito:2015fra} & $\Pi(0)$ + $\Pi(s)_{\rm{latt}}$ + $\Delta\Pi(0)$\\
\hline
-4.08 (44)	& +0.21 (34)	&	0 (5)	&	this work & $\Pi(0)$ + ${\cal C}_0^{\mbox{\scriptsize{pheno/R$\chi$T}}}$ + $1/N_c$\\
-4.17 (35)	& -0.43 (12)	&	-1 (6)	&	this work & $\Pi(0)$ + $\Delta\Pi(0)$ + $1/N_c$\\
\hline
\end{tabular}
\caption{Compilation of recent determinations of the LECs. The determinations of $\Leff$, {\it i.e.} $\Pi(0)$, are obtained as explained in Table~\ref{tab:compilation}. $1/N_c$ refers to Eq.~\eqref{Ncguess}, whereas $\Delta\Pi(0)$ refers to the sum rule given in Eq.~\eqref{fbsr}.
Additional details are given in the text.}
\label{tab:compilation-bis}
\end{center}
\end{table}

An alternative sum rule involving $L_{10}^r$ and ${\cal C}_0^r$ was recently derived in Ref.~\cite{Golterman:2014nua} from an analysis of the flavour-breaking left-right correlator $\overline\Pi^{(0+1)}_{ud-us,LR}(0)$, namely\footnote{We use the value obtained in Ref.~\cite{Golterman:2014nua} using 1999 OPAL data for the non-strange part, $0.0113\; (15)$, instead of the more precise value of Ref.~\cite{Boito:2015fra} from 2014 ALEPH data, $0.0111\; (11)$, in order to avoid possible correlations with our determination of $\Leff$.}
\beqn\label{fbsr}
\left[\overline\Pi^{(0+1)}_{ud,LR}(0) - \overline\Pi^{(0+1)}_{us,LR}(0)\right]_{\mathrm{LEC}}
&=&-0.7218\, L_5^r + 1.423\, L_9^r + 2.125\, L_ {10}^{r}
- \frac{m_{K}^{2}-m_{\pi}^{2}}{m_{\pi}^{2}}\, {\cal C}_0^{r}
\nonumber \\ &=& 0.0113\; (15)\, ,
\eeqn
again at $\mu=M_\rho$. Combining this constraint with the sum rule\footnote{We use $L_5^r(M_{\rho})=(1.19 \pm 0.25)\cdot 10^{-3} $\cite{Dowdall:2013rya} and, once again, $L_9^r (M_{\rho}) = 5.93\; (43) \cdot 10^{-3}$~\cite{Bijnens:2002hp}.}
in Eq.~\eqref{L10-p6-num} and the naive inequality in Eq.~\eqref{Ncguess}, we obtain the results shown in Table~\ref{tab:compilation-bis} (6th row) and Fig.~\ref{fig:C0C1plot} (dark blue region). We see that $L_{10}^r$ is in excellent agreement with the value obtained using Eqs. (\ref{const1}-\ref{const3}) and has a smaller error. Concerning the NNLO LECs, almost the same value is obtained for ${\cal C}_1^r$, whereas a 1.8 $\sigma$ tension is present in the ${\cal C}_0^r$ case.

Another interesting development was performed in Ref.~\cite{Boyle:2014pja}, where additional constraints on $L_{10}^r$, ${\cal C}_0^r$ and ${\cal C}_1^r$ were obtained from lattice simulations of the correlator $\overline\Pi(s)$ at unphysical meson masses. As shown in Table~\ref{tab:compilation-bis}, the lattice data allow for a more accurate determination of the LECs, making unnecessary the use of the naive guess in Eq.~\eqref{Ncguess}. However, to derive the lattice constraints one needs to assume that the ${\cal O}(p^{6})$ $\chi$PT expansion reproduces well the correlator at $s\sim-0.25\, \mathrm{GeV}^{2}$, the energy region with smaller lattice uncertainties, which dominates these constraints. Unfortunately, it was shown in Ref.~\cite{Boito:2012nt} that ${\cal O}(p^{6})$ $\chi$PT does not approximate well enough $\overline\Pi(s)$ at these energies, taking into account the low uncertainties we are dealing with, and one needs to incorporate the so-far unknown $\cO(p^8)$ chiral corrections.

In order to take advantage of the most precise lattice constraint, Ref.~\cite{Boito:2015fra} makes the strong assumption that the missing $\cO(p^8)$ chiral contributions are dominated by mass-independent terms, {\it i.e.},
$\overline{\Pi}(s)\approx\overline{\Pi}^{\chi \mathrm{PT}}_{{\cal O}(p^{6})} + D\, s^{2}$, so that they cancel in the lattice-continuum difference
$\Pi^{\chi\mathrm{PT}}_{\mathrm{lattice}}- \Pi^{\chi\mathrm{PT}}_{\mathrm{physical}}$.
It is worth noting that this is not a good approximation at the previous chiral order, $\cO(p^6)$, since more than 25\% of the $\cO(p^6)$ correction proportional to $s$ comes from known mass-dependent chiral terms.
Therefore, the uncertainties associated with these lattice constraints seem at present underestimated.

Additionally, correlations between the continuum and the lattice sum rules ({\it e.g.} due to $L_9^r$) are not publicly available. It is worth mentioning nonetheless that if we implement these lattice constraints\footnote{We find that the constraint associated to the third lattice ensemble used in \cite{Boito:2015fra} fully dominates the fits.} (instead of the inequality in Eq.~(\ref{Ncguess})), neglecting such correlations,
we reproduce the results of Ref.~\cite{Boito:2015fra} except for the uncertainties associated to $L_5^r$ and $L_9^r$, for which the neglected correlations are likely to be relevant. Such an agreement is not surprising, as our determinations of the effective coupling $\Leff$ were very close.

From Table~\ref{tab:compilation-bis} and Fig.~\ref{fig:C0C1plot} we see that the determinations obtained with the lattice constraints are (in most cases) significantly more precise than those using instead the inequality of Eq.~\eqref{Ncguess}. The agreement is reasonable (in the $0.5-1.7 \,\sigma$ range depending on the quantity), taking into account that Eq.~\eqref{Ncguess} is nothing but a naive educated guess, while the lattice improvement suffers from additional uncertainties not yet included in the quoted errors.

%
%
%
The determination of $C_{87}^r$ from $C_{87}^{\rm eff}$ at $\cO(p^6)$ does not involve any unknown LEC. The relation (\ref{C87-p6}) contains a one-loop correction
of size $-(3.16\pm 0.13)\cdot 10^{-3}$, which only depends on $L_9^r(M_\rho)$ and the
pion and kaon masses,
and small non-analytic two-loop contributions collected in the
term $G'_{2L}(M_\rho,s=0) = -0.28\cdot 10^{-3}\:\mathrm{GeV}^{-2}$. 
In spite of its $1/N_C$ suppression, the one-loop correction is
very sizeable, decreasing the final value of the $\cO(p^6)$ LEC:
\be
\label{valC87}
C_{87}^r(M_\rho)\; =\; (5.10 \pm 0.22) \cdot 10^{-3} \:\mathrm{GeV}^{-2}\, .
\qquad
\ee

\subsection{Previous determinations with other methods}

Our phenomenological determinations of $L_{10}^r(M_\rho)$ and $C_{87}^r(M_\rho)$ from $\tau$ decay data are in good agreement with the large-$N_C$ estimates based on lowest-meson dominance \cite{Knecht:2001xc,Cirigliano:2006hb,Cirigliano:2004ue,Pich:2002xy,Amoros:1999dp,Bijnens:2014lea}:
\beqn
L_{10} & = & -\frac{F_V^2}{4 M_V^2} + \frac{F_A^2}{4 M_A^2}\;\approx\; -\frac{3 f_\pi^2}{8 M_V^2}\;\approx\; -5.4\cdot 10^{-3}\, ,
\no\\
C_{87} & = & \frac{F_V^2}{8 M_V^4} - \frac{F_A^2}{8 M_A^4}\;\approx\; \frac{7 f_\pi^2}{32 M_V^4}\;\approx\; 5.3\cdot 10^{-3}\;\mathrm{GeV}^{-2}\, .
\eeqn
They also agree with the $C_{87}$ determinations based on Pade approximants~\cite{Masjuan:2007ay,Masjuan:2008fr}, which are however unable to fix the renormalization-scale dependence that is of higher-order in $1/N_C$.

The resonance chiral theory (R$\chi$T) Lagrangian \cite{Cirigliano:2006hb,Cirigliano:2004ue,Ecker:1989yg,Ecker:1988te} was used to analyze the left-right correlator at NLO in the $1/N_C$ expansion in Ref.~\cite{Pich:2008jm}. Matching the effective field theory description with the short-distance QCD behavior, both LECs are determined, keeping full control of their $\mu$ dependence. The predicted values \cite{Pich:2008jm}
\beqn
L_{10}^r(M_\rho) & = & -(4.4 \pm 0.9) \cdot10^{-3}\, ,
\no\\
C_{87}^r(M_\rho) & = &  (3.6 \pm 1.3) \cdot10^{-3}\;\mathrm{GeV}^{-2}\, ,
\eeqn
are in good agreement with our determinations, although they are less precise.

Lattice determinations of the $\chi$PT LECs have improved considerably in recent times, although they are still limited to $\cO(p^4)$ accuracy. The most recent simulations find:
\be
L_{10}^r(M_\rho)\; = \; \left\{ \bat
- (5.7 \pm 1.1\pm 0.7) \cdot 10^{-3} &  \mbox{\protect{\cite{Boyle:2009xi}}}\, ,
\\
- (5.2 \pm 0.2\, {}^{+\, 0.5}_{-\, 0.3})  \cdot 10^{-3} &  \mbox{\protect{\cite{Shintani:2008qe}}}\, .
\ea\right.
\ee
These lattice results are in good agreement with our determinations, but their accuracy is still far from the phenomenological precision.

\section{Conclusions}

We have determined the LECs $\Leff$ and $\Ceff$,  using the recently updated ALEPH spectral functions \cite{Davier:2013sfa}, with the methods developed in Refs.~\cite{GonzalezAlonso:2008rf,GonzalezAlonso:2010rn,GonzalezAlonso:2010xf}. Our final values, obtained using pinched weight functions with a
statistical analysis that includes possible DV uncertainties, are:
\beqn
\Leff&=& (-6.48 \pm 0.05) \cdot 10^{-3}\, ,
\label{eq:finalL10eff}\\[5pt]
\Ceff&=& (8.40  \pm 0.18) \cdot 10^{-3}\; \mathrm{GeV}^{-2}\, .
\label{eq:finalC87eff}\eeqn
These results are in excellent agreement with the values extracted with non-pinched weights and with those determined neglecting DV in Eqs.~(\ref{neg1}) and (\ref{neg2}). Thus, DV does not play any significant role in the determination of LECs, where the weight functions strongly suppress the high energy region of the spectral integrations. Our results are in good agreement with the ones obtained previously with the 2005 release of the ALEPH $\tau$ data  \cite{GonzalezAlonso:2010xf}:
\beqn
\Leff&=& (-6.44 \pm 0.05) \cdot 10^{-3}\, ,
\\[5pt]
\Ceff&=& (8.17  \pm 0.12) \cdot 10^{-3}\; \mathrm{GeV}^{-2} \, .
\eeqn
The improvements introduced in the 2014 ALEPH data set did not bring major changes in these parameters. The values in Eqs.~\eqn{eq:finalL10eff} and \eqn{eq:finalC87eff} also agree with the results obtained recently with the same experimental data but with a different approach in Ref.~\cite{Boito:2015fra}.

The statistical approach adopted in our analysis allows for a precise determination of the dimension-6 and 8 terms in the OPE of the left-right correlator $\Pi(s)$.
We obtain:
\beqn
\mathcal{O}_{6}&=& \; (-3.6 \,{}^{+\, 0.7}_{-\, 0.6}) \cdot 10^{-3}\; \mathrm{GeV}^{6}\, ,
\\[5pt]
\mathcal{O}_{8}&=& \; (-1.0 \pm 0.4) \cdot 10^{-2}\; \mathrm{GeV}^{8}\, ,
\eeqn
also compatible with the determinations performed in Refs.~\cite{GonzalezAlonso:2010xf} (with non-updated ALEPH data) and \cite{Boito:2015fra}
(with a different approach for estimating DV effects).
Using the same method, some higher-dimensional
terms in the OPE have also being estimated in Eqs.~(\ref{eqhigher1})-(\ref{eqhigher4}).

The numerical determination of the effective couplings $\Leff$ and $\Ceff$ has allowed us to derive the corresponding LECs of the $\chi$PT Lagrangian. At $\cO(p^6)$, we find
\beqn
L_{10}^r(M_{\rho}) & = &  -(4.1 \pm 0.4) \cdot 10^{-3}\, ,
\label{l10-final} \\
C_{87}^r(M_\rho) & = & (5.10 \pm 0.22) \cdot 10^{-3} \:\mathrm{GeV}^{-2}\, .
\label{c87-final}
\eeqn
The final value quoted for $L_{10}^r(M_{\rho})$ takes into account our two different estimates in Table~\ref{tab:compilation-bis}, keeping conservatively the individual errors in view of the present uncertainties induced by the NLO LECs.

\section*{Acknowledgments}
This work has been supported in part by the Spanish Government and ERDF funds from
the EU Commission [Grants No. FPA2014-53631-C2-1-P and FPU14/02990], by the Spanish Centro de Excelencia Severo Ochoa Programme [Grant SEV-2014-0398] and by the Generalitat Valenciana [PrometeoII/2013/007]. M.G.-A. is grateful to the LABEX Lyon Institute of Origins (ANR-10-LABX-0066) of the Universit\'e de Lyon for its financial support within the program ANR-11-IDEX- 0007 of the French government.

\appendix

\section{Low-energy expansion of the left-right correlation function}
\label{app:Pi}

At low energies, the correlator $\Pi(s)$ can be expanded in powers of momenta over the chiral symmetry-breaking scale. The series expansion has been calculated to $\cO(p^6)$
in $\chi$PT~\cite{Gasser:1984gg,Gasser:1984ux,Amoros:1999dp}:
\beqn\label{eqpi(s)}
\nonumber\Pi(s)&=&\dfrac{2f_{\pi}^{2}}{s-m_{\pi}^{2}}-8L_{10}^r-8B_{V}^{\pi\pi}(s)-4B_{V}^{KK}(s)
\\[5pt]
\nonumber&+&16\,C_{87}^r\, s\;-32\,m_{\pi}^{2}\, (C_{61}^{r}-C_{12}^{r}-C_{80}^{r})
\\[5pt]
\nonumber&-&32\,(m_{\pi}^{2}+2m_{K}^{2})\, (C_{62}^{r}-C_{13}^{r}-C_{81}^{r})
\\[5pt]
\nonumber&+&16\left((2\mu_{\pi}+\mu_{K})(L_9^r+2L_{10}^r)-
\left[2B_{V}^{\pi\pi}(s)+B_{V}^{KK}(s)\right]
\,L_9^r\,\frac{s}{f_{\pi}^{2}}  \right)
\\[5pt]
&-&8\,G_{2L}(s) \; ,
\eeqn
where
\beqn
B_{V}^{ii}(s)&\equiv& -\dfrac{1}{192\pi^{2}}\,\left(\sigma_{i}^{2}\, \left[\sigma_{i}\,\log{\left(\frac{\sigma_{i}-1}{\sigma_{i}+1}\right)}+2\right]
- \log{\left(\frac{m_{i}^{2}}{\mu^{2}}\right)}-\frac{1}{3}\right) \; ,
\\[5pt]
\sigma_{i}&=&\sqrt{1-\frac{4m_{i}^{2}}{s}}  \; ,
\\[5pt]
\mu_i&\equiv& m_i^2 \log(m_i/\mu)/(16 \pi^2 f_\pi^2) \; ,
\eeqn
and $G_{2L}(s)$ is the two-loop contribution. The analytic expression of $G_{2L}(s)$ is too large to be given here, even in the $s\rightarrow0$ limit; it can be extracted from Ref.~\cite{Amoros:1999dp}.
For $\mu=M_{\rho}$, the numerical values for its contribution and its derivative at $s=0$ are:
%
\beqn
G_{2L}(0)&=&-0.53\cdot10^{-3} \; ,
\\[5pt]
G'_{2L}(0)&=&-0.28\cdot10^{-3}\; \mathrm{GeV}^{-2} \; .
\eeqn

\bibliography{mybibfile}
\end{document}